\documentclass{optica-article}
\journal{opticajournal}
\articletype{Research Article}

\begin{document}
\renewcommand{\hbar}{\mathchar'26\mkern-9mu h}

\title{Tunable in situ Near-UV Pulses by a Transient Plasmonic Resonance in Nanocomposites}




\author{Anton Husakou,\authormark{1,*}  Ihar Babushkin,\authormark{1,2,3} Olga Fedotova,\authormark{4} Ryhor Rysetsky,\authormark{4} Tatsiana Smirnova,\authormark{5} Oleg Khasanov,\authormark{4} Alexander Fedotov,\authormark{5} Usman Sapaev,\authormark{6} Tzveta Apostolova\authormark{7,8}} \address{\authormark{1}Max Born Institute, Max Born Str. 2a, 12489 Berlin, Germany\\\authormark{2}Institute of Quantum Optics, Leibnitz Hannover University, Welfengarten 1, 30167 Hannover, Germany\\\authormark{3}Cluster of Excellence PhoenixD (Photonics, Optics, and Engineering – Innovation Across Disciplines), Welfengarten 1, 30167 Hannover, Germany\\\authormark{4} Scientific and Practical Materials Research Center, Belarus NAS, Brovky 17, 220072 Minsk, Belarus\\\authormark{5}Belarus State University, Niezalie\u{z}nasci avenue 4, 220030 Minsk, Belarus\\\authormark{6}Tashkent State Technical University, 2 uy 2 Qatartol ko'chasi, 100097 Tashkent, Uzbekistan\\\authormark{7}Institute for Nuclear Research and Nuclear Energy, Bulgarian Academy of Sciences, Tsarigradsko Chausse 72, 1784 Sofia, Bulgaria\\\authormark{8}Institute for Advanced Physical Studies, New Bulgarian University, 1618 Sofia, Bulgaria} \email{\authormark{*}gusakov@mbi-berlin.de}


\begin{abstract}
We propose a new concept for generation of ultrashort pulses based on transient plasmonic resonance in nanoparticle composites. Photoionization and free-carriers plasma change the susceptibility of nanoparticles on a few-femtosecond scale. This results in a narrow time window during the pump pulse duration when the system is in a plasmonic resonance, accompanied by a short burst of the local field. During this process, frequency-tunable few-fs pulses are generated. We elucidate the details of the above mechanism, and  investigate the influences of different contributing processes.
\end{abstract}




\section{Introduction}
Numerous fields of modern ultrafast optics, such as tracing of atomic motion in molecules \cite{zewail}, chemistry on electronic timescale \cite{remacle}, steering of ultrafast electron dynamics in the valence shell of  solids, nanoparticles, and clusters \cite{remacle2,lunnemann}, two-dimensional electronic spectroscopy experiments \cite{nenov}, generation and diagnosis of warm-dense matter \cite{chi},  material modification\cite{mod} and so on require short, sub-10-fs intense pulses at UV or near-UV frequencies. The ultraviolet (UV) wavelength range is of great interest for ultrafast spectroscopic investigations because of the possible resonance with electronic transitions of many small molecules with fairly simple excited state energy level structure, whose photo-induced dynamics can be accurately modeled using ab initio computational approaches. Transient absorption  spectroscopy in the UV can thus be used for the study of the optical response of bio-molecules and can benchmark the accuracy of such methods. 

In the visible and near infrared ranges, ultrashort pulses are routinely generated using different approaches: directly from a laser oscillator \cite{Ell}, by a non-collinear optical parametric amplifier \cite{Brida}, or by spectral broadening in a nonlinear medium (e.g. solid or hollow-core optical fiber) \cite{Nisoli} based on  self-phase modulation due to the optical Kerr-effect \cite{Khodakovskiy} or time-dependent plasma density \cite{Tempea, Apostolova}. The extension of these techniques to the UV range is challenging due to the lack of broadband laser gain media in the UV (except for excimers of noble gas halides \cite{Nagy}) and strong two-photon absorption for high-energy photons \cite{Tzankov} required for UV optical parametric amplifiers. This is why broadband UV pulses are typically generated in a two-step approach: first, few-optical-cycle pulses in the visible or near-infrared  ranges are generated, and then nonlinear frequency up-conversion is used to reach the UV range.

For frequency up-conversion in gases, the usually used nonlinear processes are third- or higher- order harmonic generation or four-wave-mixing between the fundamental wavelength and the second harmonic of the driving pulse, which is typically obtained from a Ti:sapphire laser. To name a few examples, 16-fs pulses at 266 nm were produced by third-harmonic generation in air from the 20-fs pulses \cite{Backus}, much shorter sub-4-fs pulses at 270 nm were generated in neon from the 6-fs laser output spectrally broadened in a hollow-core fiber \cite{Graf}, 8-fs pulses at 266 nm were generated by four-wave-mixing between the second harmonic and the fundamental wavelength from a 20-fs  pulse \cite{Durfee}, and 5-nJ pulses at 133 nm were generated by cascaded four-wave-mixing between third, fourth and fifth harmonic in a filament \cite{Horio}. UV pulses were also generated by second-order nonlinear frequency conversion in crystals, typically $\beta$-barium borate. For example, sub-10-fs 400-nm pulses were reported \cite{Varillas}, with sufficient spectral bandwidth obtained via frequency-doubling using broadband phase-matching enabled by grating recollimation. Furthermore, UV pulses with 9.7 fs pulse duration were generated by second harmonic generation using a spectrally shaped 1.9 mJ, 8 fs few-cycle near-infrared pulses \cite{Xiao}.

Recently, new trends appeared in the generation of UV pulses.
Resonant dispersive-wave generation based on Kerr nonlinearlity during the optical soliton propagation in waveguides combines intrinsically short pulse duration and easy tunability over the full ultraviolet spectral range as well as the entire visible spectrum when using infrared pump pulses, as was recently demonstrated \cite{Brahms,Lekosiotis}. These  advances also raise the prospect of compact on-chip integrated nonlinear devices \cite{Liu_1}. 

Due to a persistently high demand for short pulses in near-UV range and adjacent visible frequencies, it is promising and timely to search  and investigate alternative approaches to their generation, preferably directly from relatively long infrared pulses without compression step.

A possibility to generate short UV pulses by a very small device in situ would be an important and highly desirable feature, particularly in view of biological applications. Indeed, during propagation in any transparent condensed matter, a $\sim$10-fs pulse will very quickly, on the sub-mm scale, become much longer due to group-velocity dispersion. Pre-compensation of group-velocity dispersion is in principle possible but challenging and requires precise {\it a priori} knowledge of the material properties, which is rarely available. Therefore it would be highly valuable to suggest a technique which allows generation of short pulse directly inside of a transparent material at the desired position. On top of that, spectral tunability of the pulses would strongly enhance their application potential, e.g. by allowing to address different optical transitions.

\begin{figure}[htbp]
  \centering
  \includegraphics[width=0.45\textwidth]{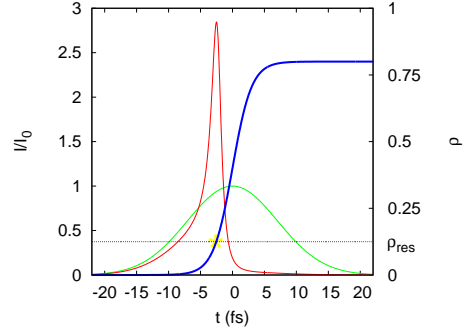}
\caption{The schematic representation of the key idea of the investigation. In a nanocomposite, the intensity inside the spherical nanoparticles (red curve) can be high compared to incident intensity (green curve) during the short time range when the plasmonic resonance (indicated by the yellow asterisk) is reached. The resonance takes place when the relative plasma density (blue curve) crosses the resonant value $\rho_{\mathrm{res}}$ indicated by the dashed black curve.}
\end{figure}

Here we propose and investigate such a method, based on a nanoparticle (NP) composite. The key idea of the paper is illustrated in Fig. 1. The incident long near-ir pulse, shown by the green curve, leads to transition of electrons from the valence zone to the first and higher conduction zones inside the dielectric nanostructure. The motion of these electrons is almost free (possibly with modified effective mass). Therefore they provide a negative Drude-type contribution to the dielectric function of the NPs $\epsilon_i(t)$, which decreases in time due to growing density of carriers $\rho$ (illustrated by blue curve in Fig. 1). The plasmonic resonance is determined by the condition  $2\epsilon_h+\epsilon_i(t)=0$, where $\epsilon_h$ is the dielectric function of host material. For $2\epsilon_h+\epsilon_i(-\infty)>0$, the plasmonic resonance can be reached only for a short time range when the relative free-carried density is close to a certain value $\rho_{\mathrm{res}}$. In this time range, the local field inside of the NPs, proportional to $1/(2\epsilon_h+\epsilon_h(t))$, shows a burst as illustrated by the red curve in Fig. 1. As will be discussed later, this burst and associated nonlinear processes lead to generation of short pulses at new frequencies (above the frequency of the pump field). All of the above aims: short near-UV pulse generation directly from long IR pulses, generation in situ at the position on NPs, as well as spectral tunability, are met by this design.   

The paper is organized as follows. In section 2, we present the physical model used for simulation of nonlinear pulse propagation. In section 3, we show the results regarding the nonlinear dynamics and short pulse generation, both in time and spectral domains. In section 4, we discuss the tunability of short pulses and the responsible mechanisms. The summary of the results is provided in the conclusion.

\section{Theoretical model}
The model used in this paper is based on the formalism described in our recent work (for details, see Ref. \cite{our_arxiv} and discussion therein), with important modifications pertinent to the time-dependent contribution of free-carrier plasma.

We consider a composite consisting of a homogeneous host material and spherical NPs, randomly distributed in space, with diameter well below the light wavelength so that effective-medium theory can be applied. The following effects are included into account: linear dispersion including intrinsic and scattering losses, second- and third-order optical nonlinearities, as well as photoionization accompanied by ionization losses and plasma dynamics. A unidirectional (1+1)D propagation equation \cite{hh,mg} is the most suitable for this kind of situations:
\begin{eqnarray}
\frac{\partial E(z,\omega)}{\partial z}&=&-i\frac{[n_{\mathrm{eff}}(\omega)-n_g(\omega_0)]\omega}{c}
E(z,\omega)\nonumber\\&&-\frac{i\omega}{2cn_{\mathrm{eff}}(\omega_0)}P_{\mathrm{NL}}(z,\omega),
\label{main_e}
\end{eqnarray}
where $E(z,\omega)$ is the Fourier transform of the electric field $E(z,t)$, $z$ is the propagation coordinate, $n_{\mathrm{eff}}(\omega)$ is the refractive index, $n_g(\omega)$ is the group refractive index, $\omega_0$ is a characteristic frequency of the pulse spectrum,  and $P_{\mathrm{NL}}(z,\omega)$ is the Fourier transform of the nonlinear part of the polarization. No slowly-varying envelope approximation is used, and $E(z,t)$ represents the real-valued field including the carrier oscillations. This approach provides a unified treatment for a pulse with an arbitrary spectral content. 

\begin{figure*}[t!]
  \centering
  \includegraphics[width=0.9\textwidth]{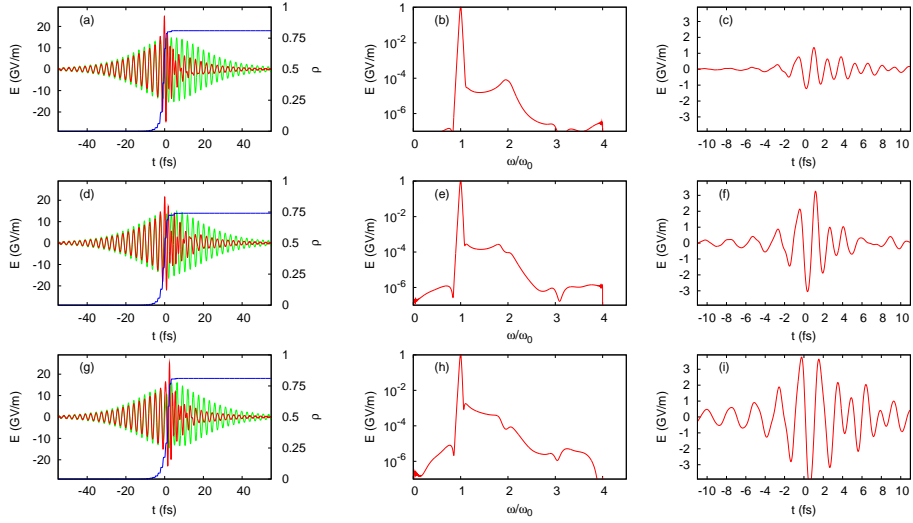}
\caption{The temporal properties [left column, (a),(d),(g)], spectra of the average field E$_{\mathrm{av}}$ [middle column, (b),(e),(h)], and temporal profiles generated short pulses [right column, (c),(f),(i)] for propagation distances of 15 nm [top row, (a),(b),(c)], 45 nm [middle row, (d),(e),(f)], 105 nm [bottom row, (g),(h),(i)]. Input pulse centered at 800 nm has a duration of 25 fs and intensity of 25 TW/cm$^2$. A composite of AlN NPs with identical radius of 2.5 nm and filling factor of $f=0.003$ in SiO$_2$ host is considered. In left column, composite-averaged electric field (green curve), local field inside the NPs (red curve), and relative plasma density (blue curve) are shown.}
\end{figure*}

The effective-medium theory allows to describe the nanocomposite material as a homogenised medium with appropriately defined effective material parameters. For low volume filling fractions of the spherical NPs $f$ and moderate scattering loss, the effective refractive index of a composite is given by \cite{mg}
\begin{equation}
n_{\mathrm{eff}}=\sqrt\epsilon_h+\frac32\frac{f(\epsilon_i-\epsilon_h)}{2\epsilon_h+\epsilon_i}+i\sqrt{\epsilon_h}{c}\left(\frac{\epsilon_h-\epsilon_i}{2\epsilon_h+\epsilon_i}\right)^2\left(\frac{r_{\mathrm{NP}}\omega}{c}\right)^3,
\label{neff}
\end{equation}
 where $r_{\mathrm{NP}}$ is the NPs radius, and $\epsilon_{h,i}$ are the dielectric functions of the host and of the NPs, correspondingly. 

We consider the situation when the presence of plasma leads to significant Drude-type modification of the dielectric susceptibility of the NPs:
\begin{equation}
\epsilon_i\to\epsilon_i-\frac{N\rho(t)e^2}{\epsilon_0m_e\omega(\omega+i\nu)},
\label{mod}
\end{equation}
where $N$ is the density of the neutral atoms or molecules before the ionization, $\rho(t)$ is the time-dependent  relative density of the plasma, $m_e$ is the effective electron mass in the conduction zone, and $\nu$ is the collision rate of the conduction electrons. Note that we consider composites with ionization rate in host material lower than that in NPs and neglect the plasma contribution to $\epsilon_h$.

One can see that the above Eq. (\ref{mod}) combines quantities defined both in frequency ($\omega$) and time ($t$) domain. This issue can be resolved by formally substituting $\omega\to i\partial_t$. Let us consider the ratio $x=E_{\mathrm{loc}}/E_{\mathrm{av}}$ between the local field inside the NPs $E_{\mathrm{loc}}$, and the average field in the composite $E_{\mathrm{av}}$. For stationary materials (i.e., those without any time-dependent parameters), it is given by:
\begin{equation}
    x(\omega)=\frac{3\epsilon_h(\omega)}{2\epsilon_h(\omega)+\epsilon_i(\omega)}.
    \label{x}
\end{equation}
In a general time-dependent case, knowing $x$ allows to directly calculate the factor $1/(2\epsilon_h(\omega)+\epsilon_i(\omega))=x/(3\epsilon_h)$ which appears in Eq. (\ref{neff}). Combining Eq. (\ref{mod}) and (\ref{x}), we obtain  
\begin{eqnarray}
    &&\left[\frac{\partial^2}{\partial t^2}+\nu\frac{\partial}{\partial t}\right]E_{\mathrm{loc}}(t)=\kappa\rho(t)E_{\mathrm{loc}}(t)+\nonumber\\&&\frac{3\epsilon_h}{2\epsilon_h+\epsilon_i}\left[\frac{\partial^2}{\partial t^2}+\nu\frac{\partial}{\partial t}\right]E_{\mathrm{av}}(t),\\
    &&\frac{\partial \rho(t)}{\partial t}=\Gamma(E_{\mathrm{loc}}(t)),
    \label{system}
\end{eqnarray}
where $\Gamma$ is the ionization rate (see more details below). We have neglected the frequency dependence of $\epsilon_h$ and $\epsilon_i$ in this term (but not in linear polarization) for pump frequency well below the bandgap. For each position in $z$, $E_{\mathrm{av}}(t)$ is a known function determined from the propagation equation (\ref{main_e}).
This system of equations is the key novel contribution of the presented model. It allows to calculate the time-dependent ratio $x=E_{\mathrm{loc}}/E_{\mathrm{av}}$ including the contribution of plasma.

The second equation in the above system describes the photoionization inside of NPs, which is induced by the local field inside of the NPs $E_{\mathrm{loc}}(t)$. Due to high relevance of the photoionization for the considered process, it is critically important to develop an accurate formalism for the ionization rate in agreement with the experiment for both of the NP materials considered in this paper: AlN and ZnO. Two models for the photoionization were considered: one based on Ivanov-Yudin formalism\cite{yi} and one based on ADK formula\cite{adk}. In femtosecond regime the damage threshold (DT) is associated\cite{dtbr} with the intensity at which the plasma contribution would lead to strong back-reflection in bulk (corresponding to bulk dielectric function near zero), which allows us to benchmark the models by the experimental data. For AlN, data regarding the DT is available for several values of the pulse duration\cite{aln1dam,aln2dam}, suggesting not the typical $\tau^{1/2}$\cite{tau05} but a $\tau^{1/4}$\cite{tau025} law for increase of DT with pulse duration. Our calculations show that while Ivanov-Yudin model provides poor agreement with the experiment, the ADK formula is surprisingly accurate in predicting the DT for a range of the pulse durations, therefore ADK photoionization rate was used, with an insignificant phenomenological prefactor of $1.35$. For ZnO, less data regarding the DT is available\cite{znodam}, therefore for ZnO we augmented the ADK rate by a first-principle calculation of the ionization rate based on
the numerical solution of the time-dependent 3D Schrodinger equation in single active electron approximation \cite{ab}. In this approach the empirical pseudopotential method was used for the electron band structure of ZnO  \cite{fan}. These calculations provided outstanding agreement with the available experimental data, with numerical threshold differing from the experimental one by 5\%. Therefore a prefactor obtained from the time-dependent 3D Schrodinger equation was utilized for the photionization rate $\Gamma(E_{\mathrm{loc}}(t))$.

The second- and third-order nonlinear processes can also be described in the framework of the effective-medium theory. The expressions for the effective second- and third-order order susceptibility in a stationary medium look like \cite{sb}
\begin{eqnarray}
    &&\chi_{\mathrm{eff}}^{(2)}(\omega_1=\omega_2+\omega_3;\omega_2,\omega_3)=(1-f)\chi_{h}^{(2)}\nonumber\\&&+fx(\omega_1)x(\omega_2)x(\omega_3)\chi_{i}^{(2)},\\
    &&\chi_{\mathrm{eff}}^{(3)}(\omega_1=\omega_2+\omega_3+\omega_4;\omega_2,\omega_3,\omega_4)=(1-f)\chi_{h}^{(3)}+\nonumber\\&&fx(\omega_1)x(\omega_2)x(\omega_3)x(\omega_4)\chi_{i}^{(3)},
    \label{chi23}
\end{eqnarray}
where $\chi_{h}^{(2)}$ and $\chi_{i}^{(2)}$ are the susceptibilities of host and NP materials, correspondingly. Note that we neglected the frequency dependence of the bulk susceptibilities of host and NPs, which is a good assumption far from bulk resonances, as well as thermal effects which happen on a picosecond time scale. For the considered dynamic case, we derive the following expressions for the second- and third-order polarizations:

\begin{eqnarray}
    P^{(2)}(t)=(1-f)\chi_{h}^{(2)}E_{\mathrm{av}}(t)^2+f\chi^{(2)}_{i}xE_{\mathrm{loc}}(t)^2,\\
    P^{(3)}(t)=(1-f)\chi_{h}^{(2)}E_{\mathrm{av}}(t)^3+f\chi^{(3)}_{i}xE_{\mathrm{loc}}(t)^3.
    \label{pol23}
\end{eqnarray}

We solve the propagation equation by an extended split-step method, whereby each of the contributions to the polarization is treated subsequently, which allows to reduce the accumulation of numerical error, using the Runge-Kutta method of the order 4. Fixed step of the grid both in time and in the propagation coordinate is used. The appearance of numerical artifacts during the propagation is monitored by tracing the total pulse energy as well as the total energy absorbed at the boundaries of the numerical time window. 

\section{Numerical results}

In Fig. 2, the numerical results for  25-fs, 25 TW/cm$^2$ pulses at 800 nm propagating in a composite of AlN particles (volume filling fraction $f=0.003$) in SiO$_2$ host are presented. We have used available Sellmeyer-type expressions to model the dispersion of both materials\cite{dispaln,dispsio2}, phenomenological values of the nonlinear susceptibilities \cite{chi2aln,chi3aln,chi3sio2}, and the bandgap of 6.01 eV for the AlN\cite{bgaln}. Note that the fluence of above pulses is below the DT for fused silica of 1 J/cm$^2$, suggesting that ionization will be predominantly happening in the AlN NPs. We note parenthetically that absence or presence of back-reflection is determined by {\it effective} refractive index, therefore for AlN NP composite (as opposed to bulk AlN) due to low filling factor $f$ even significant levels of relative ionization in NPs will not lead to back-reflection, and damage can be avoided even for $\rho\sim 1$. 

From blue curves in left column in Fig. 2, one can see that indeed high levels of relative ionization are reached during the pulse. Slightly before the maximum of the pulse, the system passes through the plasmonic resonance, which manifests itself as a sharp peak of the field inside the NPs E$_{\mathrm{loc}}$ (red curve) as compared to the average field E$_{\mathrm{av}}$ (green curve). At later stages of propagation, the input pulse is modified and depleted in the center of the pulse due to photoionization, as can be seen in Fig. 2(g). In the spectral domain, at the initial stage of the propagation a pronounced peak is formed at roughly (but not exactly) double the input frequency $\omega_0$, which later broadens and extends to higher frequencies. In the right column of Fig. 2, we show the temporal profile corresponding to this higher-frequency spectral components, by leaving only the spectral range from 1.2$\omega_0$ to 3.5$\omega_0$. It is important to note that we do not calculate the Fourier-limited pulse, rather, all the spectral phases which result from propagation are preserved. One can see that after only 35 nm of propagation, a well-isolated short pulse with a FWHM duration of 1.9 fs and weak pedestal is formed, with energy efficiency of 1.2\% (corresponding to the efficiency determined from the peak field ratio of roughly 14\%). Subsequent propagation, as illustrated in Fig. 2(i), shows further efficiency increase, which is however accompanied by longer and less regular pulse shape. 

\section{Pulse tunability and generation mechanism}

\begin{figure}[htbp]
  \centering
  \includegraphics[width=0.5\textwidth]{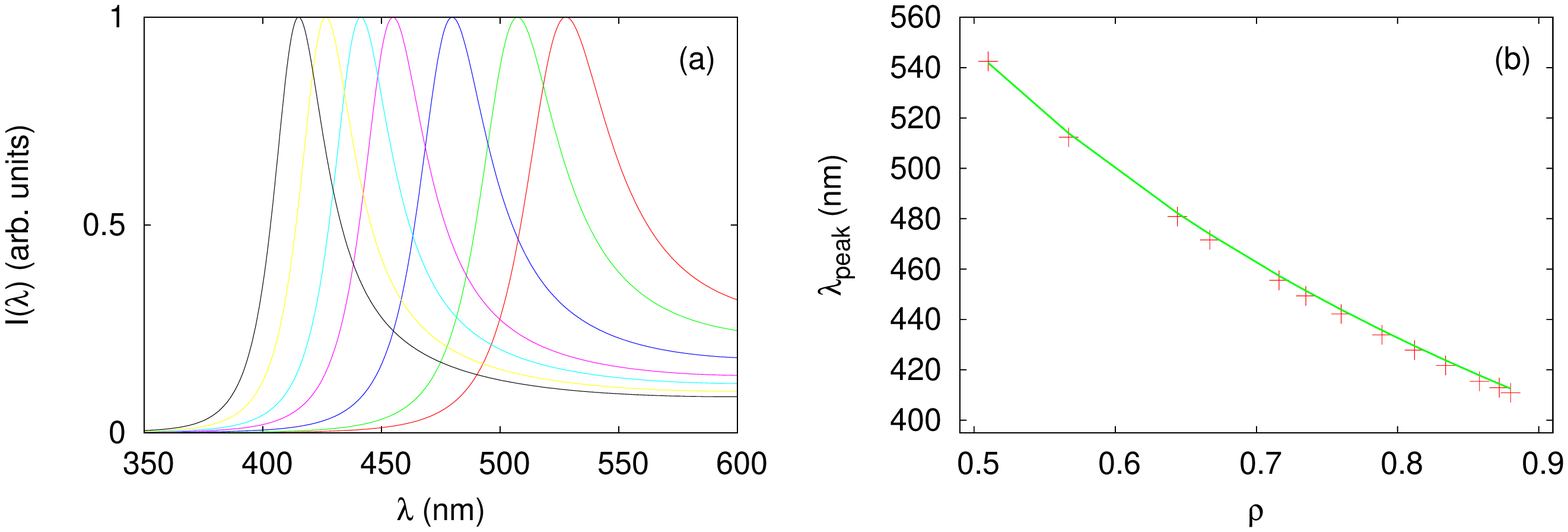}
\caption{Spectra of generated short pulse for different intensities (a) and the dependence of the central wavelength on the after-pulse plasma density (b). In (a), the intensities of 15.045 TW/cm$^2$, 15.05 TW/cm$^2$, 15.075 TW/cm$^2$, 15.2 TW/cm$^2$, 15.5 TW/cm$^2$, 17 TW/cm$^2$, 25 TW/cm$^2$ (from right to left) are considered. In (b), by solid curve the analytical dependence given by Eq. (\ref{est}) is shown.}
\end{figure}

\begin{figure}[htbp]
  \centering
  \includegraphics[width=0.45\textwidth]{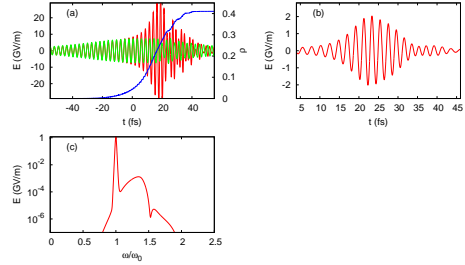}
\caption{The temporal properties (a), temporal profile of the generated short pulse (b), and spectrum (c) for propagation distance of 10 nm. Input pulse centered at 800 nm has a duration of 55 fs and intensity of 10 TW/cm$^2$. A composite of ZnO NPs with identical radius of 2.5 nm and filling factor of 0.0003 in SiO$_2$ host is considered. In (a), composite-averaged electric field (green curve), local field inside the NPs (red curve), and relative plasma density (blue curve) are shown.}
\end{figure}

We explored the possibility to influence the position of the spectral peak visible in Fig. 2(b) by varying the pump pulse intensity and, correspondingly, the relative density of plasma after
the pulse, $\rho(+\infty)$. In Fig. 3(a) we show that for different pump intensities it is possible to shift the peak (and the corresponding short pulse) in a significant spectral range, from 410 to 545 nm. In Fig. 3(b), the maximum wavelength of the peak is presented as a function of the relative density of plasma after
the pulse, $\rho(+\infty)$.

The mechanism responsible for the generation of this peak is highly relevant to understand its tunability and further features. We note that it cannot be explained by the well-known plasma-induced blue shift of the spectrum, since such shift is proportional to the nonlinear phase accumulated during propagation, and therefore the position of the peak would be $z$-dependent, in contradiction to the numerical findings. Also, the energy of the peak grow quadratically with propagation length, which excludes amplification-like processes. Rather, we speculate that the short burst at the plasmonic resonance contains many spectral components. {\it After} the plasmonic resonance, the relative plasma density $\rho(+\infty)>\rho_{\mathrm{res}}$ corresponds to the plasmonic resonance at frequency $\omega_*>\omega_0$. Spectral components at or around this frequency could be preserved and grow. To confirm this conjecture, in Fig. 3(b) we plot the wavelength corresponding to the resonant frequency after the pulse, the latter being given by

\begin{equation}
\omega_*=0.91\sqrt{\frac{Ne^2}{\epsilon_0m_e[2\epsilon_h+\epsilon_i(+\infty)]}}.
\label{est}
\end{equation}

In order to fit the numerical data, we have introduced a pre-factor of 1.21, which is justified by the fact that the peak is generated under highly dynamical conditions with plasma density quickly changing in time. In Fig. 3(b), the prediction given by Eq. (\ref{est}) is shown by the red curve. An almost perfect agreement with numerical results is obtained. Even with a fit parameter, such agreement is highly indicative that proposed mechanism indeed describes the peak generation in our system.

With the aim to investigate the influence of the material choice of the transient plasmonic resonance, in Fig. 4 we show the numerical results for the 55-fs, 10 TW/cm$^2$ pulses at 800 nm propagating in a composite of ZnO NPs (volume filling fraction $f=0.0003$) in SiO$_2$ host. Similar to the case of AlN NPs, phenomenological bulk material parameters were used\cite{disp_ZnO,sec_ZnO,thi_ZnO_1,thi_ZnO_2}. ZnO has a much lower bandgap of 3.37 eV, which has significant influence on the dynamics. The dependence of the ionization rate on the intensity is smoother and does not have a strongly pronounced threshold-like character. Therefore the growth of the relative ionization, as shown in Fig. 4(a), occurs slower, and the systems spends a longer time in the plasmonic resonance, as can be seen from comparison of local field (red curve) and average field (green curve) in Fig. 4(a). Correspondingly, the generated pulse is longer with FWHM of 9.5 fs, whereas the efficiency of 1.3\% is comparable to AlN case. Despite the quantitative differences to AlN case, the generation of the short pulse is based on the same mechanism, as can be seen from the spectrum in Fig. 4(c) showing clear isolated feature around 1.4$\omega_0$. We would like to stress that, despite the different ionization dynamics, for ZnO NPs the Eq. (\ref{est}) provides accurate estimation of the peak spectral position using the same fitting factor of 0.91. We conclude  that short pulse generation is possible for composites with host bandgap larger than inclusion bandgap, however, the latter should not be below roughly 4 eV for photoionization to be threshold-like. In addition, pedestal-free and sufficiently strong input pulses (intensity above 10 TW/cm$^2$) are required.

\section{Conclusion}

In conclusion, we have developed a model for simulation of nonlinear pulse propagation under the condition of rapid free carrier generation. We showed that a transient plasmonic resonance in a nanoparticle composite can lead to a very short burst of the local field inside the NPs. We predict a direct generation of tunable few-fs near-UV  pulses from much longer near-IR pulses, with efficiencies in the range of 1\%. The generation mechanism is connected to growth of spectral components which are in plasmonic resonance after the pulse peak. The above nonlinear dynamics is explored for two NP materials, AlN and ZnO. 

\begin{backmatter}
\bmsection{Funding}
Authors acknowledge financial support from European Union project H2020-MSCA-RISE-2018-823897 "Atlantic". I.B. thanks Cluster of Excellence PhoenixD (EXC 2122, project ID 390833453) for financial support. Support from the BNSF under Contract No. KP-06-COST/7 is acknowledged (T.A.)

\bmsection{Disclosures}
The authors declare no conflicts of interest.

\bmsection{Data availability} Data underlying the results presented in this paper are not publicly available at this time but may be obtained from the authors upon reasonable request.

\end{backmatter}

\end{document}